# Polarization switching ability dependent on multidomain topology in a uniaxial organic ferroelectric

F. Kagawa,[*,†,‡,§] S. Horiuchi,[‡,∥] N. Minami,[§] S. Ishibashi,[‡,⊥] K. Kobayashi,[¶] R. Kumai,[‡,¶] Y. Murakami,[¶] and Y. Tokura[†,§]

[†]RIKEN Center for Emergent Matter Science (CEMS), Wako 351-0198, Japan

[‡] CREST, Japan Science and Technology Agency (JST), Tokyo 102-0076, Japan

[§] Department of Applied Physics and Quantum-Phase Electronics Center (QPEC), University of Tokyo, Tokyo 113-8656, Japan

[∥] Flexible Electronics Research Center (FLEC), National Institute of Advanced Industrial Science and Technology (AIST), Tsukuba 305-8562, Japan

[⊥] Nanosystem Research Institute (NRI), National Institute of Advanced Industrial Science and Technology (AIST), Tsukuba 305-8568, Japan

[¶] Condensed Matter Research Center (CMRC) and Photon Factory, Institute of Materials Structure Science, High Energy Accelerator Research Organization (KEK), Tsukuba 305-0801, Japan

ABSTRACT:

The switching of electric polarization induced by electric fields —a fundamental functionality of ferroelectrics— is closely associated with the motions of the domain walls that separate regions with distinct polarization directions. Therefore, understanding domain-walls dynamics is of essential importance for advancing ferroelectric applications. In this Letter, we show that the topology of the multidomain structure can have an intrinsic impact on the degree of switchable polarization. Using a combination of polarization hysteresis measurements and piezoresponse force microscopy on a uniaxial organic ferroelectric, α-6,6'-dimethyl-2,2'-bipyridinium chloranilate, we found that the head-to-head (or tail-to-tail) charged domain walls are strongly pinned and thus impede the switching process; in contrast, if the charged domain walls are replaced with electrically neutral antiparallel domain walls, bulk polarization switching is achieved. Our findings suggest that manipulation of the multidomain topology can potentially control the switchable polarization.

TEXT:

Ferroelectrics exhibit spontaneous and stable polarization, and the electrically switchable nature of this polarization underlies various ferroelectric devices, such as non-volatile ferroelectric random access memory.[1] In such memory devices, the storage of data bits is achieved by driving domain walls that separate regions with different polarization directions. Ferroelectric domain walls can be classified into three types according to the relative angle between the domain-wall plane and the polarization vector $P$.[2,3] One widely observed type is electrically neutral domain walls, which have a plane parallel to $P$ (more generally, across which a normal component of *P* is continuous), whereas the other two types are positively (head-to-head) or negatively (tail-to-tail) charged domain walls, for which the plane is not parallel to *P* (more generally, across which a normal component of *P* is discontinuous) and hence has bound charges on it. If the bound charges remain electrically uncompensated, they will produce an electric field on the order of 0.1-10 MV/cm, which would exceed the polarization-switching fields of typical ferroelectrics, 1-100 kV/cm. Conversely, for charged domain walls to exist, the bound charges should be almost completely compensated by mobile charges and/or immobile charged defects.[4,5] Although their typical energies are rather high, charged domain walls have been observed in various ferroelectrics, such as $LiNbO_3$,[6] $BiFeO_3$,[7,8] lead germinate,[9] $BaTiO_3$,[10] and $Pb(Zr,Ti)O_3$ (PZT) thin films.[11]

In general, an external electric field *E* lifts the degeneracy between ferroelectric multidomains and thus exerts pressure on the domain walls to expand preferred *P* domains. However, because charged domain walls carry compensation charges, whereas neutral domain walls do not, the mobilities of the domain walls under an electric field can differ. In fact, charged domain walls are expected to be "heavy"[12,13] and to be less effectively influenced by the pressure.[14-16] Thus far, it has been observed that ferroelectric but simultaneously ferroelastic charged domain walls are strongly pinned in real space in $BiFeO_3$ thin films,[8] which is a perovskite-type multi-axial ferroelectric (*P* || <111> in the pseudocubic unit). Ferroelastic domain walls must be mechanically compatible; therefore, their possible orientations are limited.[17,18] This fact implies that the mobility of ferroelectric-ferroelastic domain walls may be affected not only by the presence or absence of compensation charges but also by the orientation-dependent elastic energy. Therefore, to verify the mobility difference between charged and neutral ferroelectric domain walls, uniaxial ferroelectrics, in which only ferroelectric but nonferroelastic domain walls are allowed, present an ideal system; however, this system remains to be explored.

In this work, we target an emergent supramolecular ferroelectric material with the hydrogen-bonded chain[19]: α-[H-6,6'-dmbp][Hca] (6,6'-dimethyl-2,2'-bipyridinium chloranilate). We demonstrate that, as a result of the mobility difference between the charged and neutral domain walls, the degree of switchable polarization strongly depends on the multidomain topology in this uniaxial organic ferroelectric. Using piezoresponse force microscopy (PFM)[20-22] and *P-E* hysteresis loop measurements, we found that multidomain states that incorporate charged domain walls do not exhibit bulk polarization switching, whereas lamella domain structures consisting of neutral domain walls do. This result is further corroborated by real-space observations of pinned charged domain walls under an electric field. Our findings suggest that by engineering the multidomain topology, the degree of switchable polarization can potentially be controlled.

The material investigated in this study is the uniaxial ferroelectric α-[H-6,6'-dmbp][Hca] (the space group is $P2_1$, and the typical sample dimensions are $2 \times 0.2 \times 0.1$ mm$^3$); hence, only ferroelectric-nonferroelastic domain walls are allowed. The schematic crystal structure at room temperature is shown in Figure 1(a), where alternating O–H⋯N and N–H⋯O hydrogen bonds form a one-dimensional chain along the *b*-axis.[23] We note that the positional order of the protons breaks the inversion symmetry and thus induces polarization along the chain (***P*** ∥ *b*), as schematically shown in Figures 1b and c. Pyroelectric current measurements confirmed that the ferroelectricity persists up to ~380 K, above which the polarization abruptly disappears (Figure 1d). Sharp anomalies associated with the ferroelectric-paraelectric transition are also detected by DSC (differential scanning calorimetry) at 378 K in a heating process and at 360 K in a cooling process (Figure 1e), and the large temperature hysteresis indicates the first-order nature of the transition.

We note that the crystal-growth temperature of this material is room temperature, which is well within the ferroelectric phase. Thus, the as-grown ferroelectric domain structure can be dominated by the crystal-growth kinetics. In contrast, when the sample temperature is increased above $T_c$ (~380 K) and then reduced to room temperature, the domain formation process is no longer relevant to the crystal-growth kinetics; therefore, the domain structures in the annealed state can differ in topology from those in the as-grown state. This unique crystal-growth situation makes α-[H-6,6'-dmbp][Hca] suitable for studying the relationship between the multidomain topology and the degree of switchable polarization, in addition to the mobility difference between charged and neutral domain walls.

To evaluate the magnitude of switchable polarization potentially exhibited by α-[H-6,6'-dmbp][Hca], first-principles calculations based on the Berry phase formalism[24,25] were performed. We introduced the parameter λ to describe the intermediate crystal structure between the reference paraelectric state (λ = 0) and the room-temperature ferroelectric state (λ = 1) (see also Methods). Intermediate structures (0 < λ < 1) were constructed through linear interpolation of the atomic positions. The spontaneous polarization was calculated by increasing λ from zero to one, and the value of 9.94 μC/cm$^2$ was obtained (Figure 2a).

Experimentally, however, much less polarization switching, ~1.3 μC/cm$^2$, was observed in the virgin *P-E* hysteresis loop at room temperature for the as-grown state (Figure 2b). For the annealed state, in contrast, we found that the switchable polarization was enhanced by greater than 500 %, thus reaching ~7 μC/cm$^2$, which is the greatest value ever reported among acid-base organic ferroelectrics.[19] This value is compatible with the result of the first-principles calculation, 9.94 μC/cm$^2$, thereby indicating bulk polarization switching for the annealed state. The slight discrepancy may be partly explained by the fact that our calculations do not incorporate finite-temperature effects.

To provide a microscopic insight into the large change in switchable polarization, we conducted in-plane PFM for the as-grown and annealed states. The as-grown topography and domain structure are shown in Figures 3a and 3b, respectively (see also Figure 4b, which shows results for the as-grown state of a different sample). The ferroelectric domain boundaries are rugged and are composed of both charged and neutral domain walls. In the annealed state, the surface morphology was degraded to some extent (Figure 3c), but an even more dramatic change can be observed in the domain structure (Figure 3d): the thermal cycling resulted in fine lamella structures that consist of neutral domain walls (note that the scales differ by approximately one order of magnitude between Figures 3b and d). All PFM images obtained for the annealed state exhibited essentially the same features as Figure 3d, which led us to conclude that the domain structure is exclusively comprised of neutral domain walls and that the degree of switchable polarization is closely linked to the multidomain topology.

The significant increase in the switchable polarization can be explained by the working hypothesis in which the charged domain walls play a resistive role in the switching process. To verify this scenario, in-situ PFM was conducted on the as-grown state before and after an in-plane electric field was applied (∥*P*) through the side electrodes (Figure 4a). To this end, we chose an area in which neutral and charged domain walls can be observed in the same view (Figure 4b). Figure 4c shows the domain state after the application of an electric field of ~21.9 kV/cm for 4 seconds, for which a lateral shift of the neutral domain walls (highlighted by dotted arrows) and the forward growth of fine domains can be clearly observed. The fine domains, which grew in a forward manner in the first stage (Figure 4c), then expanded through lateral shifts of the neutral domain walls under a stronger electric field (~27 kV/cm for 10 seconds) (Figure 4d). We note that during these processes, the pre-existing charged domain walls were strongly pinned. Although not all neutral domain walls exhibit such lateral shifts, the same propensity was also confirmed for other crystals: charged domain walls were pinned, whereas some neutral domain walls exhibited lateral shifts. These findings demonstrate that the charged domain walls are less mobile than the neutral walls under an electric field, thereby elucidating why the switchable polarization is closely associated with the multidomain topology.

The remaining issue to be discussed is why thermal cycling removes the charged domain walls that exist in the as-grown sate. It appears feasible to explain this behavior by assuming that the charged domain walls are compensated by mobile charges (Figure 5a): although the origin of mobile charges is not yet clear, their existence is substantiated by the finite conductivity of the as-grown crystal (~$10^{-13}$ $\Omega^{-1}$cm$^{-1}$). Once the system enters the paraelectric phase, the charged domain walls and their bound charges disappear. If the compensation charges still resided at the same position, they would produce a large internal electric field on the order of 1 MV/cm (Figure 5b). Therefore, the accumulated compensation charges also disappear to minimize the electrostatic energy (Figure 5c): Note that a similar redistribution of compensation charges is often observed at the polar surfaces of ferroelectrics when the magnitude of polarization changes rapidly, for example, by heating.[26,27] When the sample again enters the ferroelectric phase upon cooling through the first-order phase transition, the mobile charges cannot follow the sudden formation of charged domain walls; thus, it is difficult for charged domain walls to form. Consequently, the most stable domain structure in terms of electrostatic energy, i.e., lamella domains that consist of neutral domain walls, is preferentially formed. Obviously, this scenario does not include any material details and would be particularly relevant for ferroelectrics in which the crystal grows below Curie temperature and multiple charged domain walls are present in the as-grown state.

In conclusion, we conducted *P-E* hysteresis loop measurements and PFM on the uniaxial organic ferroelectric α-[H-6,6'-dmbp][Hca] and found that ferroelectric-nonferroelastic charged domain walls tend to be strongly pinned. Thus, the topology of the multidomain structure is an important factor that determines the polarization switching capability. The domain-wall-dependent mobility revealed in this study is important for "domain wall nanoelectronics" that exploit functionalities that emerge at the domain wall.

**Methods.** *Sample preparation*. [H-6,6'-dmbp][Hca] displays at least two polymorphisms, and we targeted a polymorph (α-form) for which the crystal structure has been previously reported.[23] The α-form, which consists of dark red elongated-plate crystals, was obtained through repetitive recrystallizations from an acetonitrile solution of a 1:1 stoichiometric mixture of purified $H_2$ca and 6,6'-dmbp.

*First-principles calculations.* The spontaneous polarization was calculated with the first-principles computational code QMAS.[28] By using experimental crystallographic data[23] and then imposing an inversion symmetry, a reference paraelectric structure was constructed. Because x-ray diffraction measurements tend to underestimate the C-H bond lengths, we exploited the target ferroelectric structure (λ = 1) for which the hydrogen positions were computationally optimized.

*PFM measurements.* PFM was conducted with a commercially available scanning probe microscope (Asylum MFP-3D). To achieve a good signal-to-noise ratio, we employed the dual-frequency resonance-tracking technique,[29] which enables imaging of domain structures in hydrogen-bonded organic ferroelectrics.[30,31]

AUTHOR INFORMATION

**Corresponding Author**

*E-mail: fumitaka.kagawa@riken.jp

**Author Contributions**

F.K. conducted the PFM imaging. S.H. grew the single crystals used for this study and conducted the DSC measurements. S.I. performed the first-principles calculations. S.H. and N.M. measured the P-E hysteresis loop. N.M. conducted the pyroelectric current measurements. S.H. and F.K. planned and led the project. F.K. wrote the article with assistance from S.H., S.I., and Y.T. All authors commented on the paper.

**Notes**

The authors declare no competing financial interest.

ACKNOWLEDGMENT

This work was partially supported by a Grant-in-Aid for Scientific Research (Grant Nos. 24224009 and 24684020) from the Japan Society for the Promotion of Science and the ``Funding Program for World-Leading Innovative R&D on Science and Technology (FIRST Program)''. This work has been performed partially under the approval of the Photon Factory Program Advisory Committee (Proposal No. 2012G115).

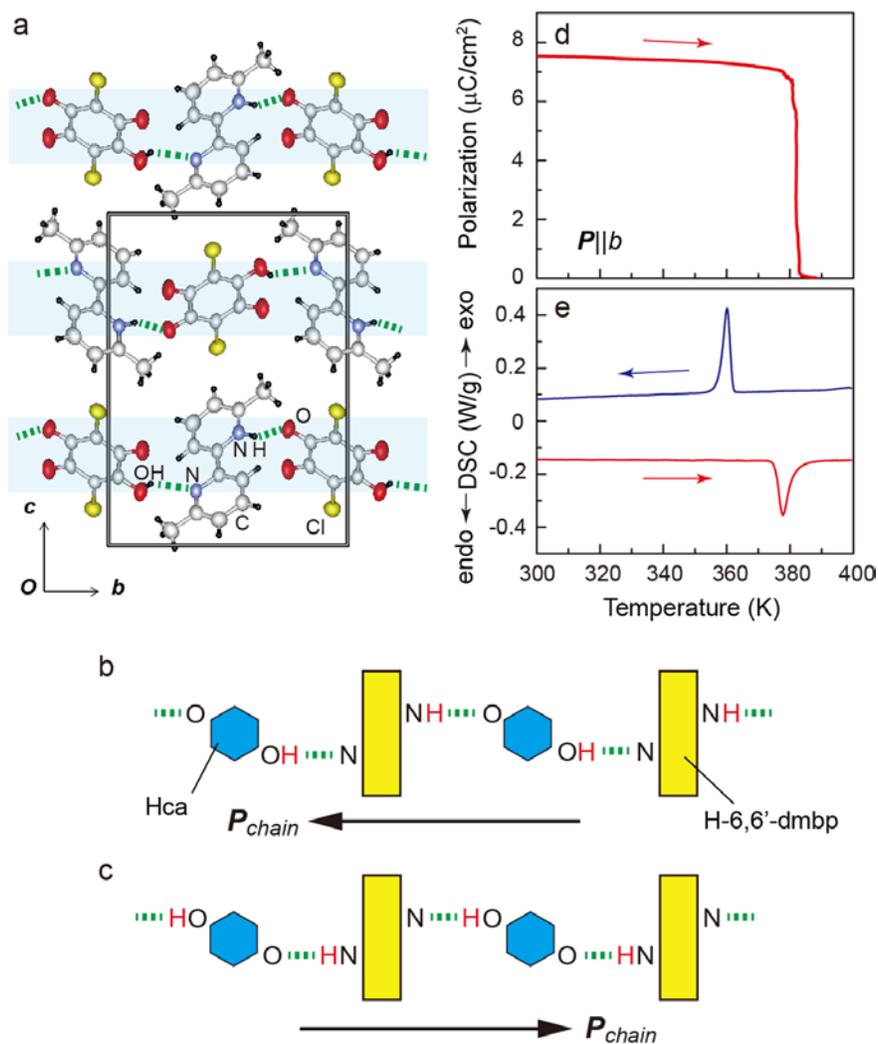

**Figure 1.** The crystal structure of α-[H-6,6'-dmbp][Hca] and the ferroelectric transition. (a) Ferroelectric α-[H-6,6'-dmbp][Hca] crystal viewed along the crystallographic *a*-axis.[23] (b and c) Schematics that illustrate the relationship between the proton position and the polarization direction of the hydrogen-bonded chain. (d) Temperature dependence of spontaneous polarization derived from pyroelectric current measurements. (e) Differential scanning calorimetry in heating and cooling processes. The dotted lines in (a)-(c) represent hydrogen bonds.

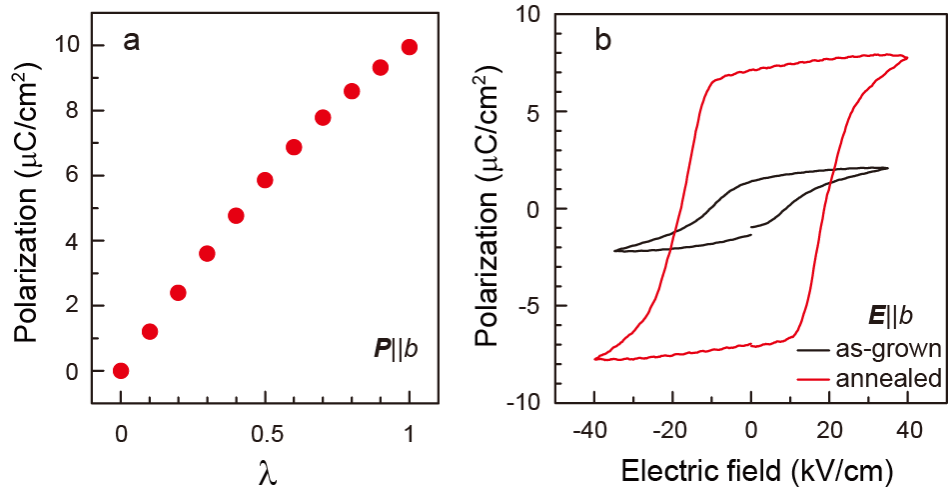

**Figure 2**. Ferroelectric properties of α-[H-6,6'-dmbp][Hca]. (a) First-principles calculations of ferroelectric polarization along the path connecting the paraelectric state ($\lambda = 0$) and the room-temperature ferroelectric state ($\lambda = 1$). (b) Polarization hysteresis curves at room temperature for the as-grown state and the annealed state.

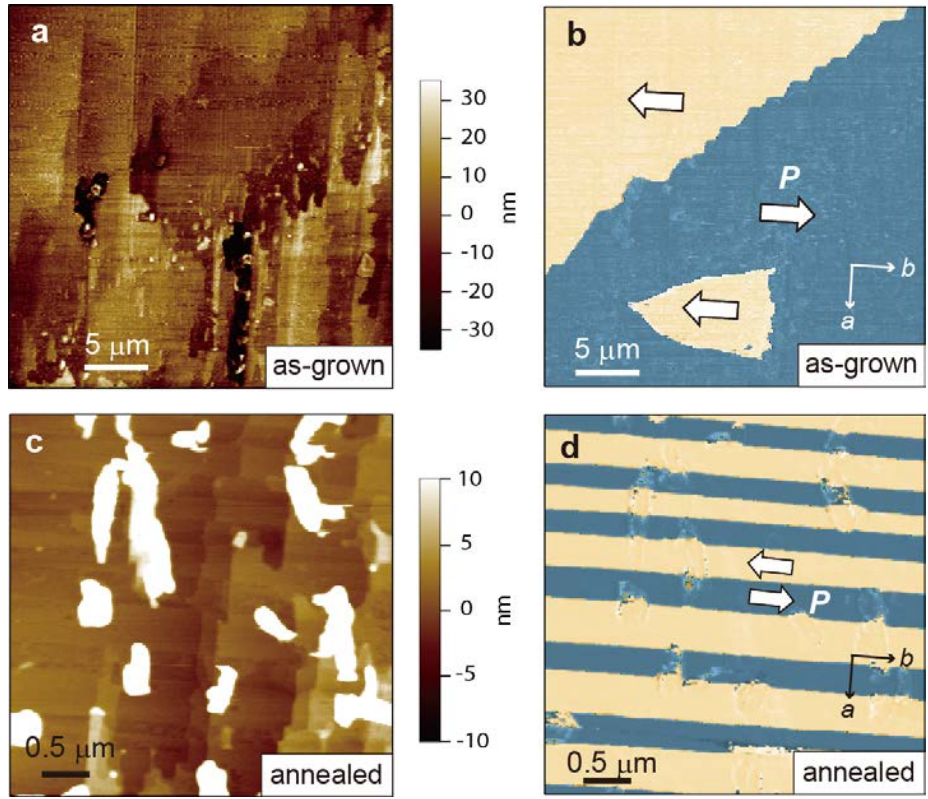

**Figure 3**. Dependence of ferroelectric domain structure on thermal history. (a, c) Surface topography and (b, d) in-plane PFM phase images in the *ab* plane at room temperature. (a) and (b) show images for the as-grown state, whereas (c) and (d) display images for the annealed state.

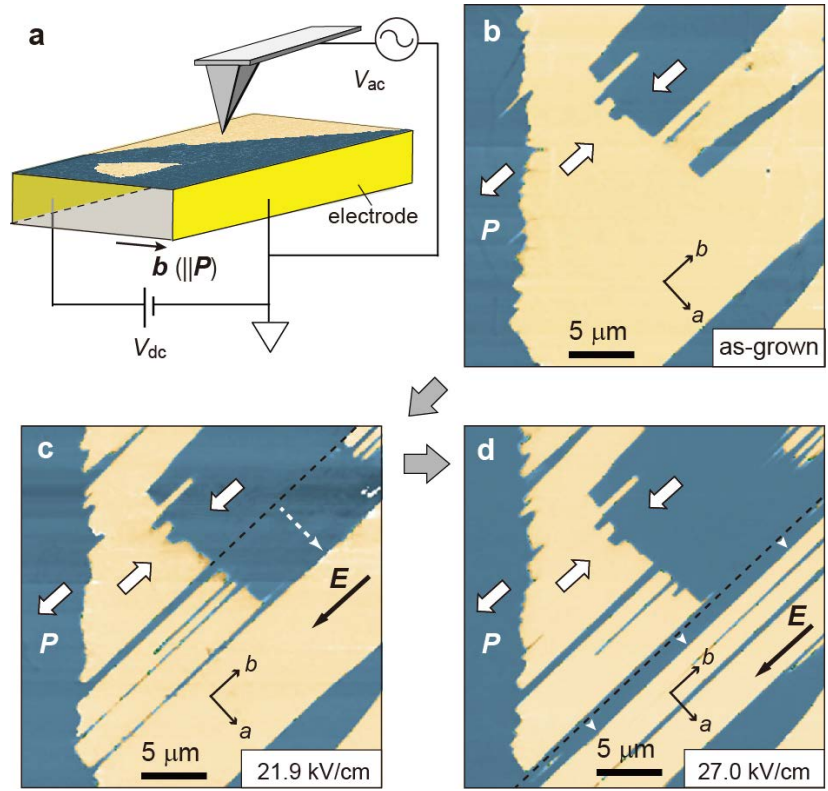

**Figure 4**. In-plane polarization switching process as observed by PFM phase images. (a) Experimental setup. (b) As-grown domain structure before the application of an electric field. (c, d) Domain structures after the application of an in-plane electric field of 21.9 kV/cm for 4 seconds (c) and 27.0 kV/cm for 10 seconds (d). The dotted lines and arrows in (c) and (d) highlight lateral shifts of neutral domain walls.

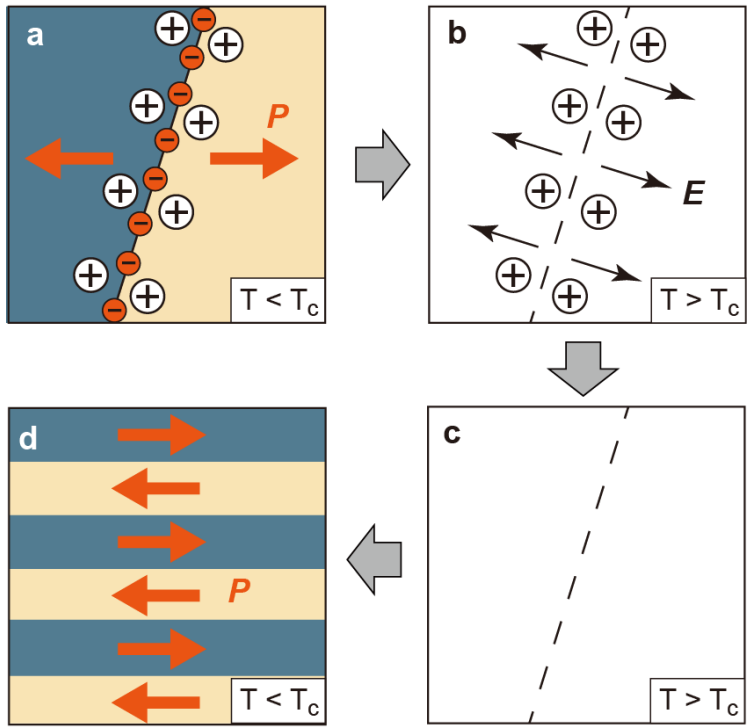

**Figure 5**. Schematic illustrations that display the domain-structure change caused by annealing. (a) As-grown domain structure, (b) paraelectric state immediately after the paraelectric phase is reached, (c) equilibrium paraelectric state, and (d) room-temperature domain structure after the paraelectric phase is experienced. The positive and negative charges represent mobile compensation charges and bound charges on the domain wall, respectively.